\documentclass[aps,prb,twocolumn,groupedaddress]{revtex4}
\usepackage{graphicx}
\usepackage{dcolumn}
\usepackage{bm}
\usepackage{amsmath,amsfonts,amssymb,amsthm,verbatim}
\usepackage[ansinew]{inputenc}
\usepackage[usenames]{pstricks}
\usepackage{color}
\usepackage{epsfig}
\usepackage{MnSymbol}
\usepackage{bbold}
\usepackage{siunitx}
\usepackage[hidelinks]{hyperref}

\newcommand{\be}{\begin{equation}}
\newcommand{\ee}{\end{equation}}
\newcommand{\bea}{\begin{eqnarray}}
\newcommand{\eea}{\end{eqnarray}}

\newcommand{\bi}{\begin{itemize}}
\newcommand{\ei}{\end{itemize}}
\newcommand{\bc}{\begin{center}}
\newcommand{\ec}{\end{center}}

\begin{document}

\title{Unsupervised learning using topological data augmentation}

\author{Oleksandr Balabanov}
\affiliation{Department of Physics, University of Gothenburg, SE 412 96 Gothenburg, Sweden}

\author{Mats Granath}
\affiliation{Department of Physics, University of Gothenburg, SE 412 96 Gothenburg, Sweden}

\begin{abstract}

Unsupervised machine learning is a cornerstone of artificial intelligence as it provides algorithms capable of learning tasks, such as classification of data, without explicit human assistance.  
We present an unsupervised deep learning protocol for finding topological indices of quantum systems. The core of the proposed scheme is a `topological data augmentation' procedure that uses seed objects to generate ensembles of topologically equivalent data. Such data, assigned with dummy labels, can then be used to train a neural network classifier for sorting arbitrary objects into topological equivalence classes. Our protocol is explicitly illustrated on 2-band insulators in 1d and 2d, characterized by a winding number and a Chern number respectively. By using the augmentation technique also in the classification step we can achieve accuracy arbitrarily close to $100\%$ even for objects with indices outside the training regime.    
\end{abstract}

\maketitle

\section{Introduction}

Machine learning (ML) techniques have shown tremendous success in completing challenging tasks in physics. The applications range from solving trademark problems in quantum physics\cite{Dunjko} to pursuing completely novel objectives. In a recent breakthrough, for example, Tshitoyan {\em et al.} [\onlinecite{Tshitoyan}]  succeeded in discovering new materials with outstanding properties by training a ML algorithm on papers published in materials science. In another recent work, Iten {\em et al.} [\onlinecite{Iten}] have introduced a specially designed encoder-decoder ML architecture, named SciNet, shown to be capable of learning  from scratch various concepts in physics. Very promising results were presented by Raccuglia {\em et al.} [\onlinecite{Raccuglia}] and Melnikov {\em et~al.} [\onlinecite{Melnikov}] where state-of-the-art ML tools have been successfully implemented for designing new experiments. {\em Quantum} algorithms for ML\cite{Biamonte} is also a topic of great current interest fueled by recent advancements in quantum computing\cite{google_quantum}. Artificial Neural Networks (NNs), one of the most efficient and widely used tools of ML\cite{LeCun}, are currently at the frontier of research activity. It has been shown that they are capable of predicting phase transitions and critical temperatures\cite{Ohtsuki, Carrasquilla, Zhang3, Chng, Beach, Rem, Casert, Kharkov, Zhang5, Huembeli1, Huembeli2}, learning topological indices of quantum phases\cite{Zhang, Zhang2, Carvalho, Caio, Ming, Mano, Wu2}, efficiently representing many-body states\cite{Carleo, Deng, Gao, Deng2, Nomura, Kaubruegger, Pastori, Levine}, improving known numerical computational methods~\cite{Bukov, Broecker, Shen}, and decoding topological quantum correcting codes\cite{Torlai,Fosel,Beenakker,sweke,andreasson}.

\begin{figure}[t] \centering
    \includegraphics[width=8.5cm,angle=0]{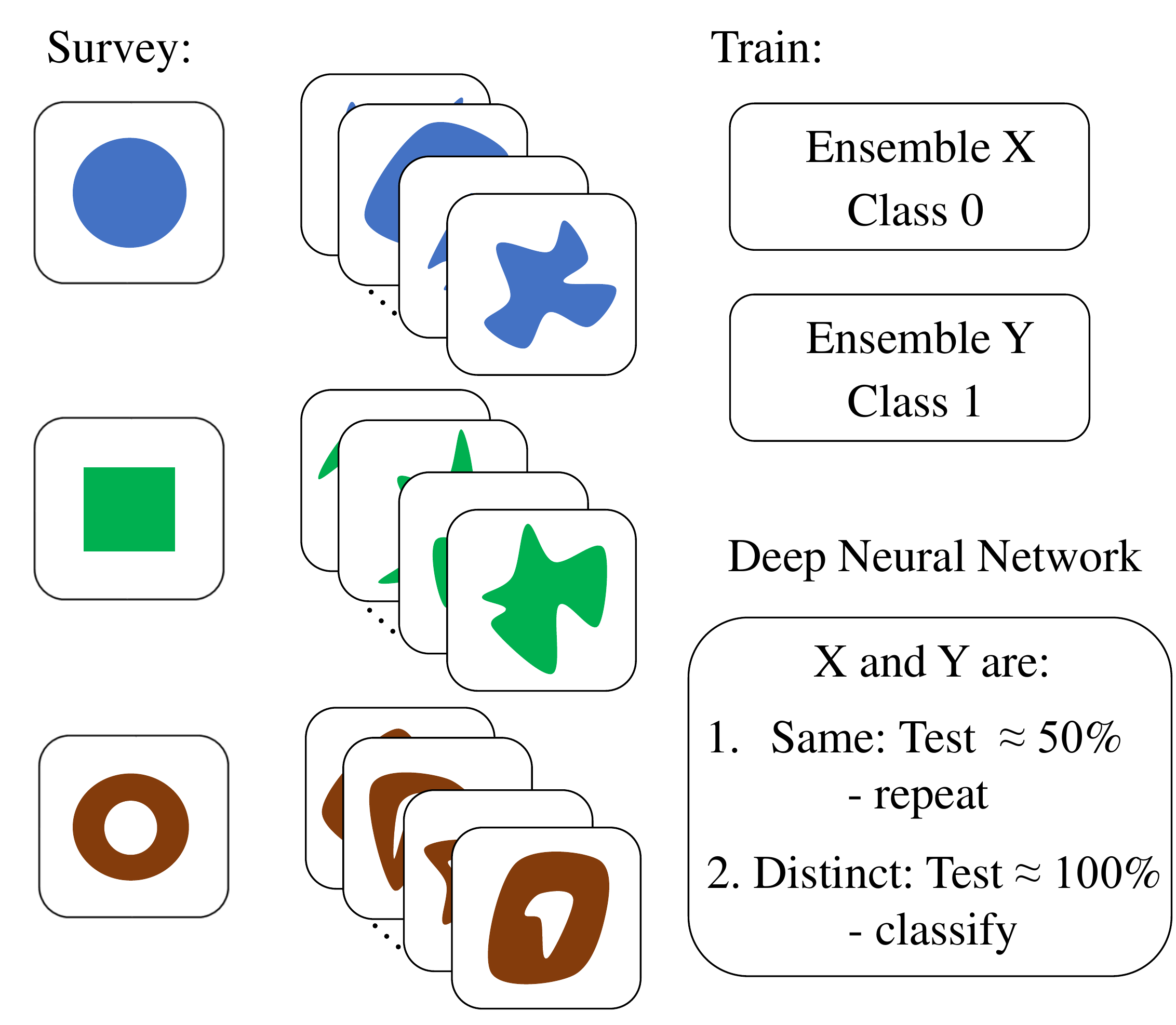}
    \caption{A schematic illustration of the `topological data augmentation' protocol for spotting topological equivalences, applied to three geometric objects of different genus in 2d. By augmenting a parent object we create topologically equivalent ensembles and use them as input data to the `learning by confusion' method by Van Nieuwenburg {\em et al.} [\onlinecite{Nieuwenburg}]. A network trained on distinct esembles can then be used for further classification of unknown objects.
    }
     \label{fig1}
\end{figure}

Neural-network-based approaches typically require training on manually labeled data before they become capable of predicting relevant features. This explicit human supervision limits  their  applicability. 
Unsupervised methods  are more scarce but have the potential to go beyond the boundaries of current knowledge. By now several unsupervised ML tools have been put forward such as usage of principal component analysis\cite{Wang, Hu, Wetzel, Costa}, variational autoencoders \cite{Hu, Wetzel, Alexandrou}, self-organizing maps \cite{Shirinyan}, advanced clustering algorithms~\cite{Rodriguez-Nieva, Durr},  divergence-based predictive methods\cite{Schafer, Greplova},  and learning by confusion~\cite{Nieuwenburg}. In this paper we introduce an unsupervised learning scheme for topological classification that takes inspiration from data augmentation techniques widely applied in image recognition \cite{Perez}. The idea, illustrated in Fig.\ \ref{fig1}, is to use a seed object, a 'parent', to generate an ensemble of 'children' that by construction are in the same category, and use such ensembles to train a neural network to categorize arbitrary objects.

Topologically nontrivial phases of matter has been a major research area in condensed matter physics for several years\cite{Ryu, Wen}. To find analytic expressions for topological indices as well as putting these in a form that allows for efficient numerical calculation can be a challenging task. The algorithm introduced in this paper bypasses the need for a closed mathematical expression for the index.  
We demonstrate this by classifying band insulators in symmetry class AIII in 1d and in symmetry class A in 2d characterized by a winding number and Chern number respectively. The algorithm essentially consists of three steps: 1) take two band insulators, one trivial and one hypothesized to be topologically non-trivial. Deform these parent objects into two large ensembles of children by means of random topology preserving transformations. 
2) Label these ensembles '0' and '1' and train a neural network to classify them. Using the learning by confusion scheme\cite{Nieuwenburg}, exemplified in Fig.~\ref{fig1}, failure to classify the two ensembles implies that the parent objects are not topologically distinct, thus going back to step 1 with a new ansatz for topologically non-trivial system. 3) The neural networks that are used (Figs.\ \ref{fig:1D_network} and \ref{fig:2D_network}) are specifically designed to learn and aggregate discretized local quantities in momentum space. Once successfully trained on a trivial and non-trivial ensemble we find that these networks can accurately classify any band insulator in the relevant symmetry class by its topological index, as exemplified in Figs.\ \ref{fig:random_class1D} and \ref{fig:random_class2D}. (See also, Figs.\ \ref{fig:random_random_1D} and \ref{fig:random_random_2D}.) 

The algorithm is bona fide `unsupervised learning'; relying only on having a formalism for performing continuous, topology preserving, deformations on a discretized representation of the object. This is the crucial main distinction from the important work by Zhang {\em et\ al.} [\onlinecite{Zhang}] and Sun {\em et al.\ } [\onlinecite{Zhang2}] that also use neural networks to calculate the topological index of band insulators. The latter rely on having access to an auxiliary independent calculator of the index in order to generate the ensembles of labeled training data. The idea developed in the present paper  removes this auxiliary calculator, thus opening up for the possibility to quantify topological phases beyond those with a known expression for the index.

The paper is organized as follows. In Sec. \ref{sec:1D} we describe our protocol applied to the topological classification of 2-band insulators in 1d and the extraction of winding numbers, which is then extended to 2d and Chern numbers in Sec.\ \ref{sec:2D}. A short summary of the main concepts and outlook for future developments are given in Sec.\ \ref{sec:summary}. In appendices we discuss training with unknown Chern number and give a detailed derivation of continuity criterion for the 2d case.   

\section{Methods \& Results}

\subsection{Band insulators in 1D, Winding number\label{sec:1D}}

As our first illustration, we implement the protocol for 1d topological classification in symmetry class AIII of the standard ten-fold classification, known to contain topologically inequivalent band insulators labeled by an integer topological invariant, the so-called winding number~$\omega$~\cite{Ryu}. Any gapped 2-band system from this symmetry class can be represented by a momentum-periodic Bloch Hamiltonian $H(k)  = h_x(k) \sigma_x + h_y(k) \sigma_y$ with some continuous functions $h_x(k)$ and $h_y(k)$, and Pauli matrices $\sigma_x$ and $\sigma_y$. The winding number $\omega$ then calculates how many times the vector $\vec{h}=(h_x, h_y)$ winds around zero as a function of $k \in [0, 2\pi)$. For efficiency it is practical to normalize $\vec{h}$ to be of unit length for each $k$ and consider from now on the space of normalized $H(k)$. 
In this formulation, any 1d gapped quantum system from symmetry class AIII can be represented by a continuous set of unit vectors $\vec{h}(k)$ with $k \in [0, 2\pi)$.

\subsubsection{Generating training data}

To demonstrate the algorithm we first study two parent systems, shown in Fig.~2, with known winding numbers $\omega=0$ and $\omega=1$. Subsequently (Sec.\ \ref{unknown_wind}) we will repeat the same analysis starting from a randomly generated parent system, where $\omega$ is unknown. Although not strictly necessary it is convenient to use as baseline a reference system that is a priori known to be trivial with $\omega=0$, here we use $H(k) = \sigma_x$. 
Based on these parent systems we create three ensembles consisting of topologically equivalent Bloch Hamiltonians and classify them by confusion:  We train a neural network to distinguish the two datasets from the trivial ensemble and if it fails/succeeds in doing so, corresponding to classification accuracy around $50 \%$/$100\%$, then the parent systems are topologically trivial/non-trivial. 

\begin{figure} \centering
    \includegraphics[width=8.5cm,angle=0]{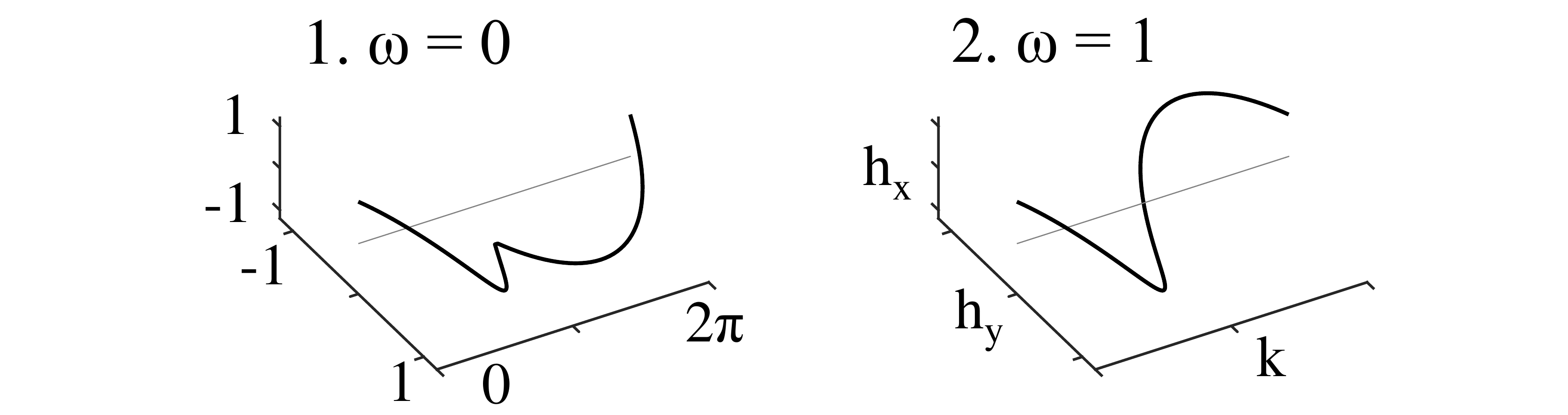} 
    \caption{Examples of 1d Bloch Hamiltonians $H(k)$ from symmetry class AIII:  1. $h_x(k) = \cos(k)$ and $h_y(k) = \sin(k)$ for $k \in [0, \pi)$ and $h_x(k) = \cos(-k)$ and $h_y(k) = \sin(-k)$ for $k \in [\pi, 2\pi)$  with $\omega = 0$; 2. $h_x(k) = \cos(k)$ and $h_y(k) = \sin(k)$ for $k \in [0, 2\pi)$ with  $\omega = 1$.}
     \label{fig2}
\end{figure}

For doing numerics we need to discretize the momentum. To rigorously define the notion of topological equivalence for discretized Bloch Hamiltonians we uniquely interpolate them to continuous ones and only then examine how they transform under the corresponding deformations. To conclude whether the change to be performed is continuous we track the interpolated objects and analytically verify that at each step the proposed change stays continuous. In this way we rigorously establish topological equivalence between the interpolated Bloch Hamiltonians before and after the deformations. In practice the discretization of momentum space is done by representing a 1d band insulator by $\vec{h}_i$ for $i\in [1,N_k]$, with $N_k$ the number of discrete momenta. To each of the discretized systems we then uniquely identify a continuous one by linearly interpolating the angle $\theta_i=\arctan(h_y/h_x)_i\in [-\pi, \pi)$. The children are produced by repeating the following deformation procedure: Randomly select a site $i$ and rotate the corresponding unit vector by a random angle $\phi \in [-\pi, \pi)$, allowing only rotations satisfying  $|\theta_{i}-\theta_{i\pm 1}+\phi| < \pi$ for avoiding any discontinuous changes.

For producing numerical results 
each of the parent systems was deformed $10^5$ times to generate one child system, collecting $10^4$ children for each ensemble of topologically equivalent systems. 
Starting with $N_k=100$ for the deformations, to alleviate the classification we additionally expand the data strings to $N_k = 200$ by linear interpolation to make sure that the relative angles $\in [-\pi, \pi)$ between any neighboring vectors are sufficiently far from the boundaries $\pm \pi$. Finally, we feed the network $201 \times 2$ layered data corresponding to vectors $(h_x, h_y)$ defined on $201$ momentum sites, where the additional $201$st vector is just a duplicate of the $1$st one, introduced to encode periodicity of the momentum space in the input.

Note that, in contrast to the original `learning by confusion' protocol~\cite{Nieuwenburg}, we do not rely on a parametrized Hamiltonian to generate data, but only on being able to form ensembles based on the topology preserving data augmentation procedure.

\begin{figure} \centering
    \includegraphics[width=8.5cm,angle=0]{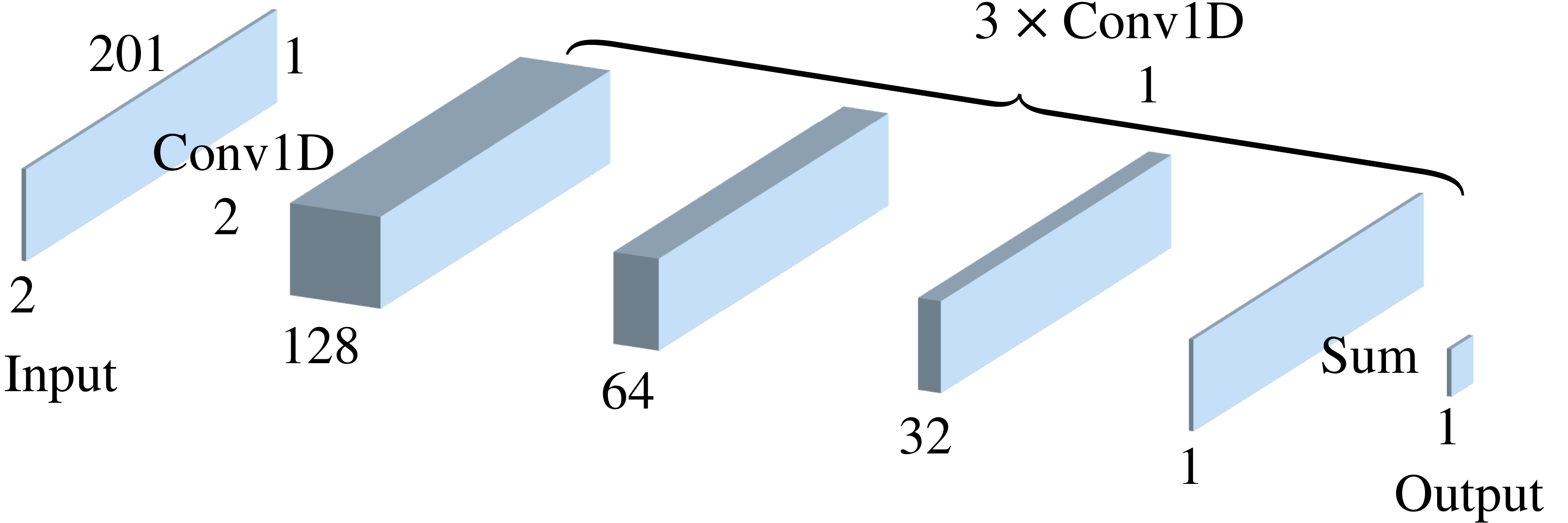}
    \caption{The neural network employed by us for classifying 1d band insulators. It takes in data of size $201 \times 2$ and consists of 1 $\times$ Conv1D layer of 128 feature maps with receptive field of size~2,  3 $\times$ Conv1D layers of 64, 32, 1 feature maps with receptive fields of size 1, and a summation layer.}
     \label{fig3}\label{fig:1D_network}
\end{figure}

\subsubsection{Neural network structure and training} 
We employ the convolutional neural network depicted in Fig. 3. It consists of a 1d convolution layer of 128 filter with receptive field of size 2, which outputs 128 feature maps as lists of size 200.  
Following this are three 1d convolution layers of 64, 32, and 1 filters with receptive field of size 1. These layers perform identical local operations on each site. Finally a summation layer gives the output in terms of a single real number corresponding to the index. The activation is rectified linear in every layer except the last convolution layer which has a linear activation function. 

The network structure is explicitly designed to sum up the output of a translationally invariant non-linear operation on each pair of neighboring sites,  thus  effectively utilizing the assumption that the topological index can be written as an integral over some unknown local function of the classified object and its derivatives. This is in contrast to the neural networks used for supervised learning of topological indices in Refs.\ \onlinecite{Zhang} and \onlinecite{Zhang2} where the networks had to learn to integrate the local features. Our network structure provides efficiency by limiting the number of trainable parameters, and by learning to calculate the relevant local quantities (see Fig.\ \ref{fig:winding_local}) it generalizes outside the training regime with high accuracy. 

In total there are 11,009 trainable parameters. The cost function is mean absolute error (L1) and the labels are chosen to be '$0$' for the reference trivial class and '$1$' for the ensemble corresponding to the parent systems from Fig.~2. We effectively augment the datasets by rotating each child system in the ensembles by some uniform random angle before each epoch. The training is done on $2 \times 9500$ samples using Adam optimizer with minibatch size $512$ and tested on the remaining $2 \times 500$ samples after each epoch. 

\subsubsection{Results}
\begin{figure} \centering
    \includegraphics[width=8.5cm,angle=0]{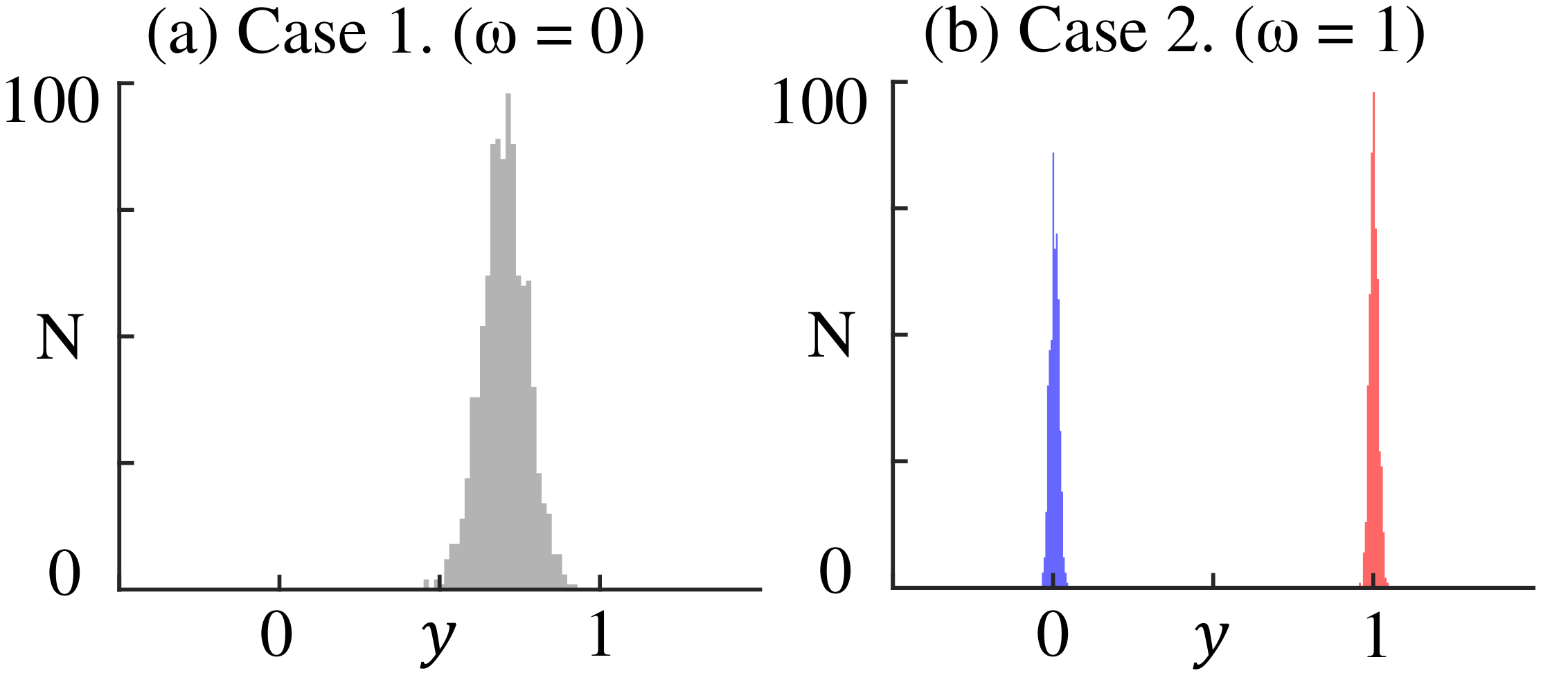} 
    \caption{A histogram plot of the network's output ($y$) evaluated on a test dataset. The network is trained using a trivial reference ensemble labeled '0' and an ensemble labeled '1' derived by data augmentation from the parent Hamiltonians 1 (a) and 2 (b) from Fig.\ \ref{fig2}. In (b) the data corresponding to the trivial (nontrivial) ensemble is illustrated in blue (red).}
     \label{fig4}
\end{figure}

In Fig.\ \ref{fig4} we plot the network's output $y$ evaluated on the test dataset after training for $500$ epochs with learning rate $10^{-4}$ and for $500$ epochs with learning rate $10^{-5}$. The results are in agreement with our expectation: The network successfully learned to distinguish the topologically nontrivial ensemble from the trivial reference, Fig.~4b, but failed in learning to differentiate the datasets of topologically trivial systems, Fig.\ \ref{fig4}a. The classification is done by taking the closest integer to the network's output $y$. The obtained classification accuracy is then $50 \%$ ($100 \%$) for the topologically (non)trivial case. The  distribution in Fig.\ \ref{fig4}a depends on initial conditions of the network but due to the L1 norm the exact position between 0 and 1 is irrelevant, with mean error always close to $0.5$.

\begin{figure} \centering
    \includegraphics[width=8.5cm,angle=0]{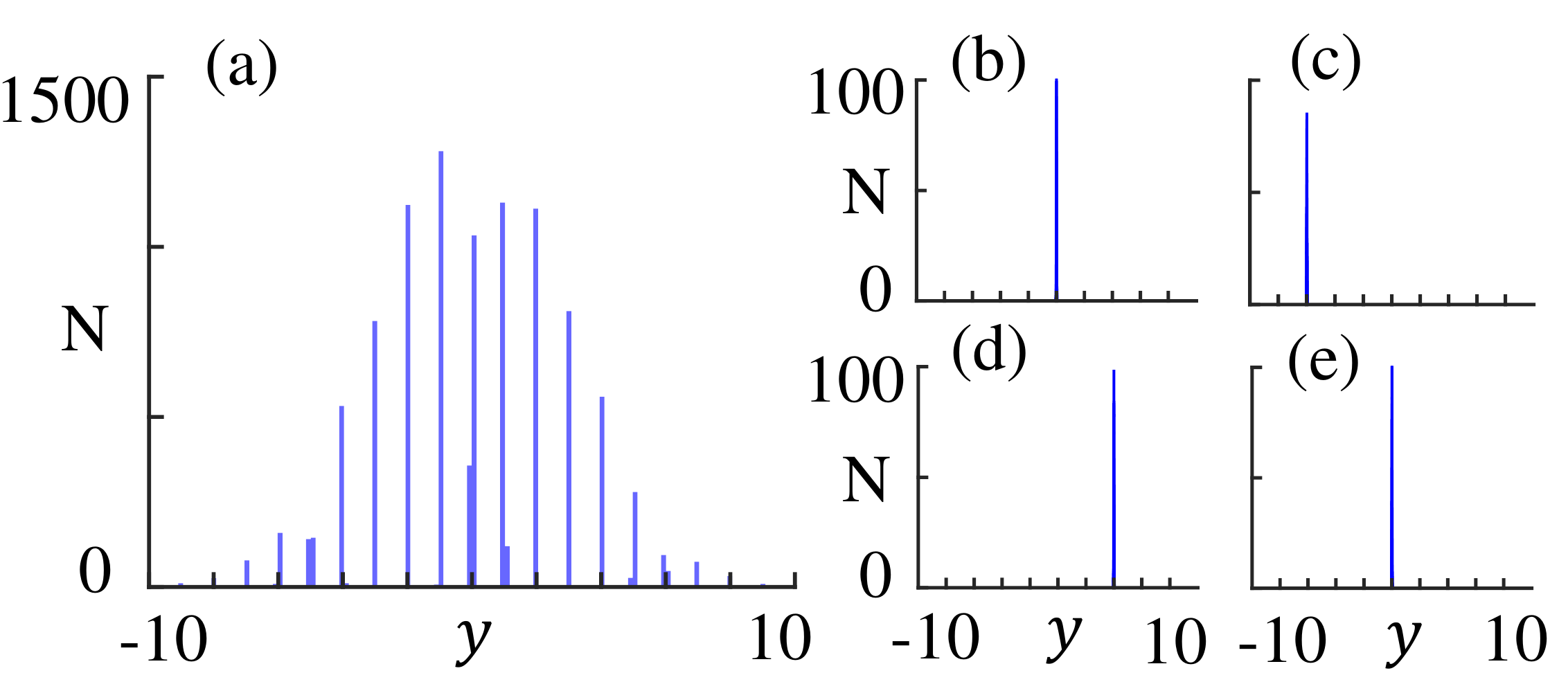}
    \caption{ The number of input child systems~N vs. network's output $y$ corresponding to (a) a dataset of $10^4$ random Bloch Hamiltonians, (b) - (e) datasets of $10^3$ samples generated from four randomly chosen parent systems. }
     \label{fig6}\label{fig:random_class1D}
\end{figure}

Quite surprisingly the network that was trained to recognize winding numbers 0 and 1 (Case 2) has actually learned to classify any object in the same class according to its topological index. To show this we generate a dataset composed of $10^4$ Bloch Hamiltonians of arbitrary winding number by taking random vectors $\vec{h}$ at each of $N_k = 100$ momentum sites, interpolated to $N_k = 200$ as for the training data. The output of the network for these objects is shown in Fig.\ \ref{fig:random_class1D}a, showing classification of the objects into bins with close to integer output. The normal distribution of values reflects the fact that the objects were generated stochastically. Note that no information about the actual winding numbers of the objects goes into this process, the raw output from the network is plotted. The network, trained with labels 0 and 1, thus generates new labels for arbitrary objects in topological sectors it has never seen, showing the power of the convolutional structure. 

To further verify that objects are actually classified correctly according to topological classes we pick four arbitrary objects from the above ensemble of stochastic Hamiltonians and use these as parent objects for generating four ensembles of topologically equivalent children using the deformation procedure (used to generate training data above). In Fig.\ \ref{fig:random_class1D}b-e we show the output from the network for these ensembles, which are indeed grouped into a single bin, corresponding to a unique winding number.

\subsubsection{Training with unknown winding number \label{unknown_wind}} 
\begin{figure} \centering
    \includegraphics[width=8.5cm,angle=0]{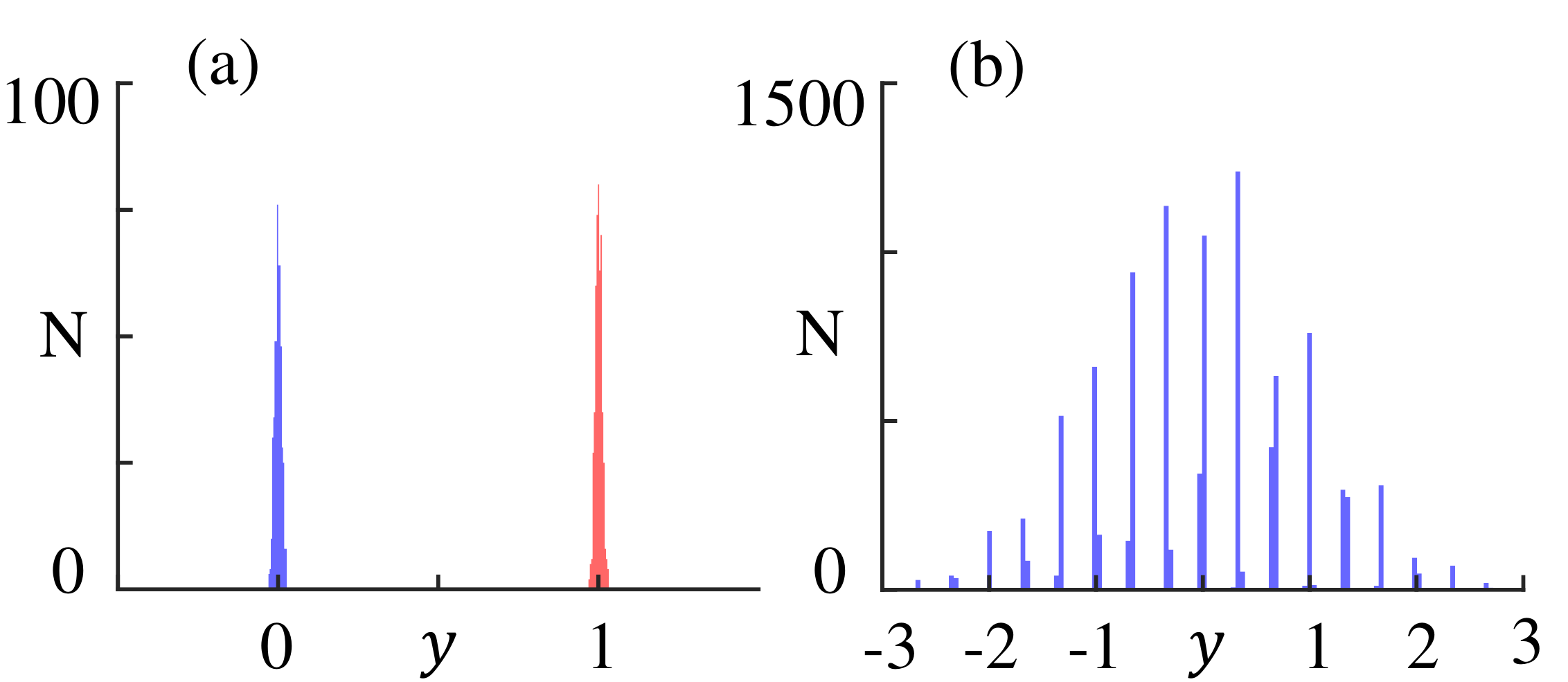}
    \caption{The outcome of our protocol performed on a randomly selected 1d parent system: The number of input systems N vs. the network's output $y$ evaluated on a) the test set with the data corresponding to the trivial reference (random parent system) depicted in blue (red), and b) a dataset of $10^4$ random 1d Bloch Hamiltonians.  }
     \label{fig_w3}\label{fig:random_random_1D}
\end{figure}

So far we used a trivial object, labeled '0' and an object with known winding number $\omega=1$, labeled '1', as parents to generate the training data, with the topological index generated by the network being identical to the actual winding number for arbitrary objects. 
We now show that it is in fact not necessary for the classification to know the winding number of the non-trivial parent to get the topological classification. Instead we can use a random parent Hamiltonian with a priori unknown winding number to generate the training ensemble with label '1'. This is exemplified in Fig.\ \ref{fig:random_random_1D} for a parent which a posteriori can be identified as $\omega=3$.  As shown in Fig.~\ref{fig:random_random_1D}b, applied to a random set of data, the network classifies objects by an integer fraction index $p/q$,  with $q\in Z$ ($3$ in this example) corresponding to the actual winding number of the non-trivial parent object. In this way we confirm that the network indeed has learned to classify 1d Bloch Hamiltonians in symmetry class AIII by their topological equivalence classes without using any prior information about the winding numbers. (We assume that we can always find a trivial '0' reference.) 

\subsubsection{Network operation}
\begin{figure} \centering
    \includegraphics[width=7.5cm,angle=0]{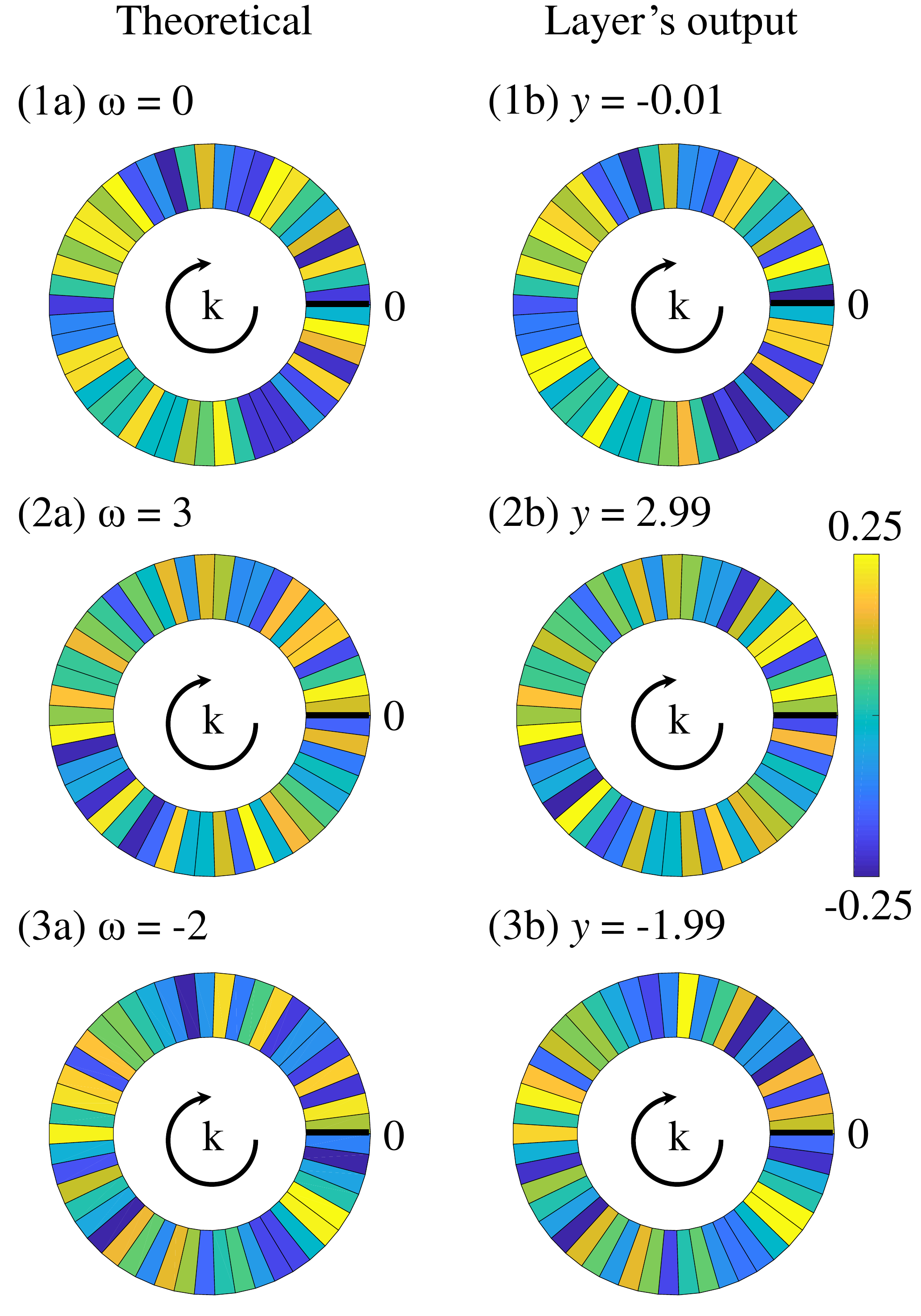}
    \caption{Samples of feature maps from the last convolutional layer  of the network in Fig.\ \ref{fig:1D_network} (right), compared to the explicitly calculated difference in neighboring angles $\Delta\theta$ (left). These are summed up to give the network output "index" and the exact winding number respectively. The close resemblance shows that the network has learned to calculate the relevant local quantity with high precision. (For visual clarity, only every fourth of 200 $k$-values are shown.) }
     \label{fig9_2}\label{fig:winding_local}
\end{figure}

To understand why the network output label is so close to the actual winding number even for objects outside the training regime we examine the last feature map which is simply summed to give the label. A few examples, corresponding to random Hamiltonians, are shown in Fig.~\ref{fig:winding_local}, where the network state is compared to a direct calculation of the discrete derivative of the angle defined by $\vec{h}(k)$. The fact that the network by itself arrives at a representation which is practically identical to the compact mathematical expression is quite striking, and related to the fact that we constrain the network to use translationally invariant units that are added up without post-processing. 

\subsection{Band insulators in 2D, Chern number\label{sec:2D}}

For second illustration we consider 2d band insulators in symmetry class~A. It is known that the topological quantum phases in this class are labeled by an integer invariant, the so-called Chern number $C\,$~\cite{Ryu}. Any 2-band insulator in symmetry class A can be written in the following form
\begin{align} 
\begin{split} 
H(\vec{k})  = \vec{h}(\vec{k})\cdot\vec{\sigma},
\end{split}
\label{eq:SSH}
\end{align}
where $\sigma_i$, $i=x,y,z$, are the Pauli matrices and $h_i(\vec{k})$ are continuous functions of 2d momentum $\vec{k}=(k_x, k_y)$, $k_x, k_y \in [0, 2\pi)$. For the purpose of studying the topological properties we can assume the vectors $\vec{h}$ to be of unit length. It follows that any 2d system in symmetry class~A can be represented by a map $\vec{h}(\vec{k})$ from 2d torus to a 3d sphere. The Chern number~$C$ then calculates how many times $\vec{h}$ wraps around the 3d sphere. 

\begin{figure} \centering
    \includegraphics[width=7.5cm,angle=0]{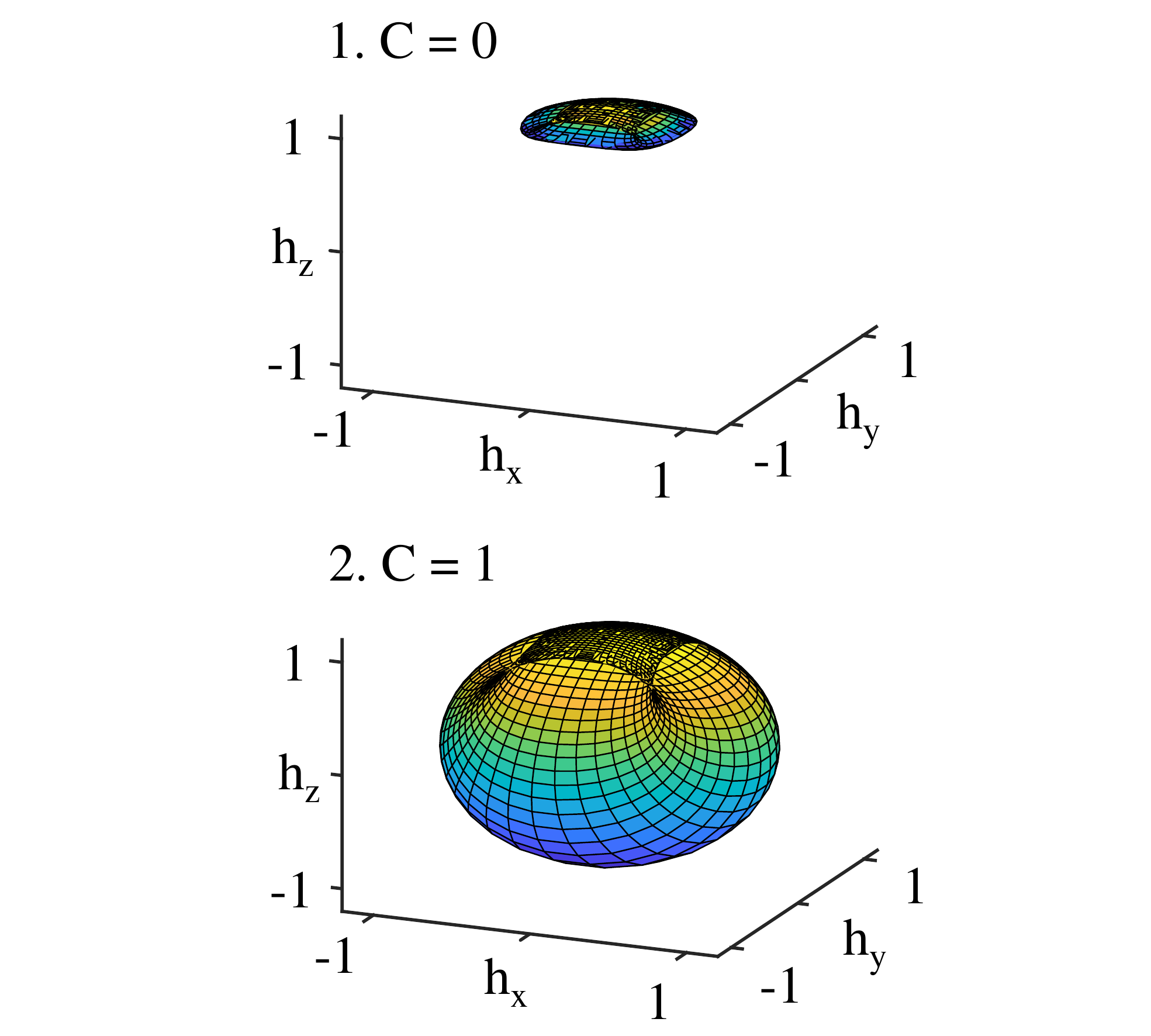}
    \caption{Examples of 2d band insulators: 1. $h_x = \sin(k_x)$, $h_y = \sin(k_y)$, $h_z  = 3 + \cos(k_x) + \cos(k_y)$ for $k \in [0, 2\pi)$ with $C = 0\,$; 2. $h_x = \sin(k_x)$, $h_y  = \sin(k_y)$, $h_z = 1.5 + \cos(k_x) + \cos(k_y)$ for $k \in [0, 2\pi)$ with $C = 1\,$.}
     \label{fig7}\label{fig:Chern}
\end{figure}

\subsubsection{Generating training data \label{sec:2D_training}}

In analogy to Sec.\ \ref{sec:1D} we here develop a scheme for finding topologically nontrivial band insulators belonging to symmetry class A in 2d. First we pick some parent representatives of this symmetry class, one topologically trivial ($C = 0$) and one nontrivial ($C \neq 0$), such as those exemplified in Fig.\ \ref{fig:Chern}. For each of the parent band insulators we then create ensembles of child systems belonging to the parent equivalence class and train a network to distinguish these datasets from a trivial reference ensemble. For the role of a trivial reference we take a topologically flat system $H = \sigma_x$ and create the corresponding ensemble of child Bloch Hamiltonians based on this parent.

\begin{figure} \centering
    \includegraphics[width=8.5cm,angle=0]{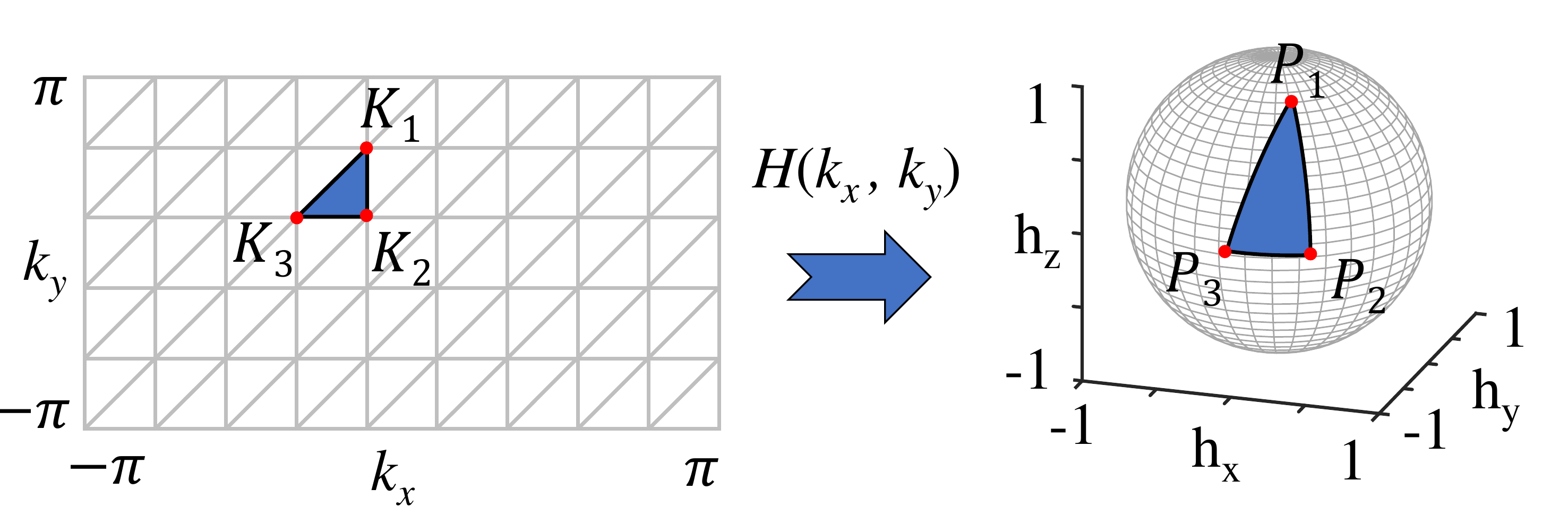}
    \caption{A correspondence between a triangle of the discretized momentum space and a triangle on a 3d sphere used for the interpolation. }
     \label{fig8}\label{fig:mapping}
\end{figure}

For implementing our algorithm we first discretize 2d momentum space by $N_x = 10$ and $N_y = 10$ equally spaced sites. The discretized Bloch Hamiltonians are put in unique correspondence with continuous ones via an interpolation procedure schematically illustrated in Fig.\ \ref{fig:mapping}. In short, we define a triangular grid in 2d momentum space and interpolate discretized $H(k_x, k_y)$ by mapping each grid triangle $(K_1, \, K_2, \, K_3)$ to the corresponding triangle $(P_1, \,P_2, \,P_3)$, with $P_i= H(K_i)$, on a sphere while satisfying the following conditions: 1. The sides of the spherical triangle are the shortest arcs connecting the vertices; 2. The face of the triangle is chosen to be the smallest spherical area enclosed by the sides. The parent objects are then deformed to generate an ensemble of topologically equivalent children. The deformations are implemented by randomly picking a site in 2d momentum space and rotating the corresponding unit vector by a random angle around the z-axis. As detailed in Appendix \ref{App:2d_transform}, we only allow those rotations that keep the corresponding continuous systems in the same topological equivalence class. To generate a single child we successively rotate $N=20$ sites in this fashion, followed by a rotation of the whole system by some new randomly chosen angles $\phi_{x}, \phi_{y}, \phi_{z} \in [-\pi,\pi)$ around all three axes, and repeat this $M=50$ times. As for the 1d problem, Sec. \ref{sec:1D}, in the very end of the deformation procedure we linearly interpolate between the momentum points, to enhance the resolution, 
giving $N_x = 20$ and $N_y = 20$. Each child system is thus represented by a 3d unit vectors defined on a $21$ by $21$ grid (extended for periodic boundary conditions), which is fed as a three channel input to the network. In total, we collect $10^4$ child Bloch Hamiltonians corresponding to each of the three parent systems. 

\subsubsection{2d Convolutional Network}
\begin{figure} \centering
    \includegraphics[width=8.5cm,angle=0]{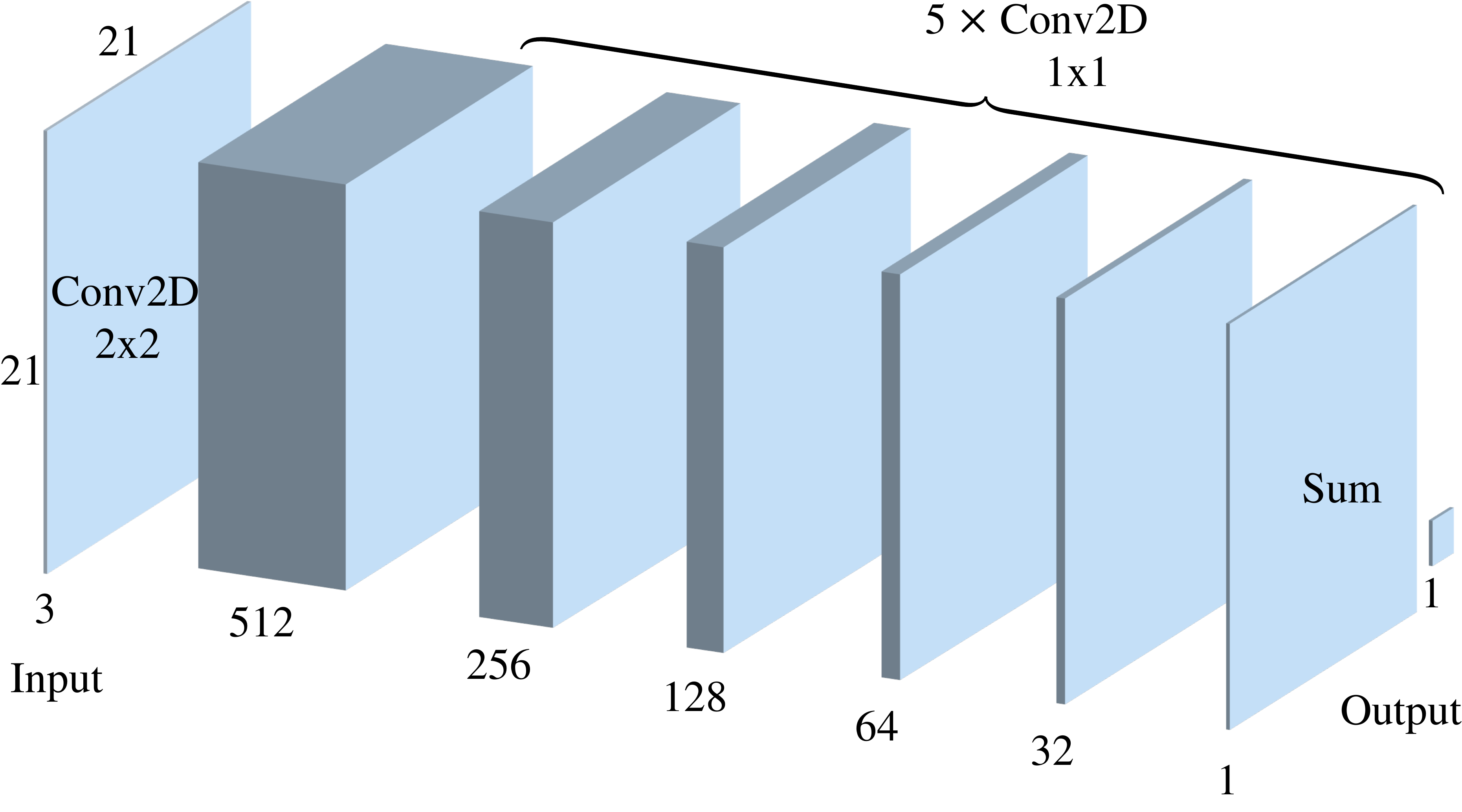}
    \caption{The neural network used for classifying topological insulators in 2d: 1 $\times$ Conv2D of 512 feature maps with $2 \times 2$ receptive field, 5 $\times$ Conv2D of 256, 128, 64, 32, 1 feature maps with $1 \times 1$ receptive fields, and a summation layer. The input is of size $21 \times 21 \times 3$.}
     \label{fig_net2}\label{fig:2D_network}
\end{figure}

The neural network trained to classify this data is similar to that in Sec.\ \ref{sec:1D} but now extended to 2d layered input as shown in Fig.\ \ref{fig:2D_network}. We employ a 2d convolutional layer of 512 filters with $2 \times 2$ receptive field, that outputs 512 feature maps of dimension $20\times 20$. The output is processed locally by means of five 2d convolutional layers of 256, 128, 64, 32, and 1 filters with $1 \times 1$ receptive field, and a summation layer. In total the network has 181,249 trainable parameters. We use rectified linear units as activation fucnctions in all trainable layers except the last convolutional layer where a linear activation is employed. The training is performed on the trivial reference ensemble (labeled '$0$') and on one of the ensembles corresponding to the parent systems from Fig.~\ref{fig:Chern} (labeled~'$1$'). We train exploiting mean absolute error cost function and Adam optimization method on minibatches of size $512$. We also effectively augment the data before each epoch by rotating every child system in the ensembles by random angles around all three axes. As before, we stress that our neural network's architecture is designed with the knowledge that the toplogical index can be represented as an integral over some unknown local quantity.

\subsubsection{Results}
\begin{figure} \centering
    \includegraphics[width=8.5cm,angle=0]{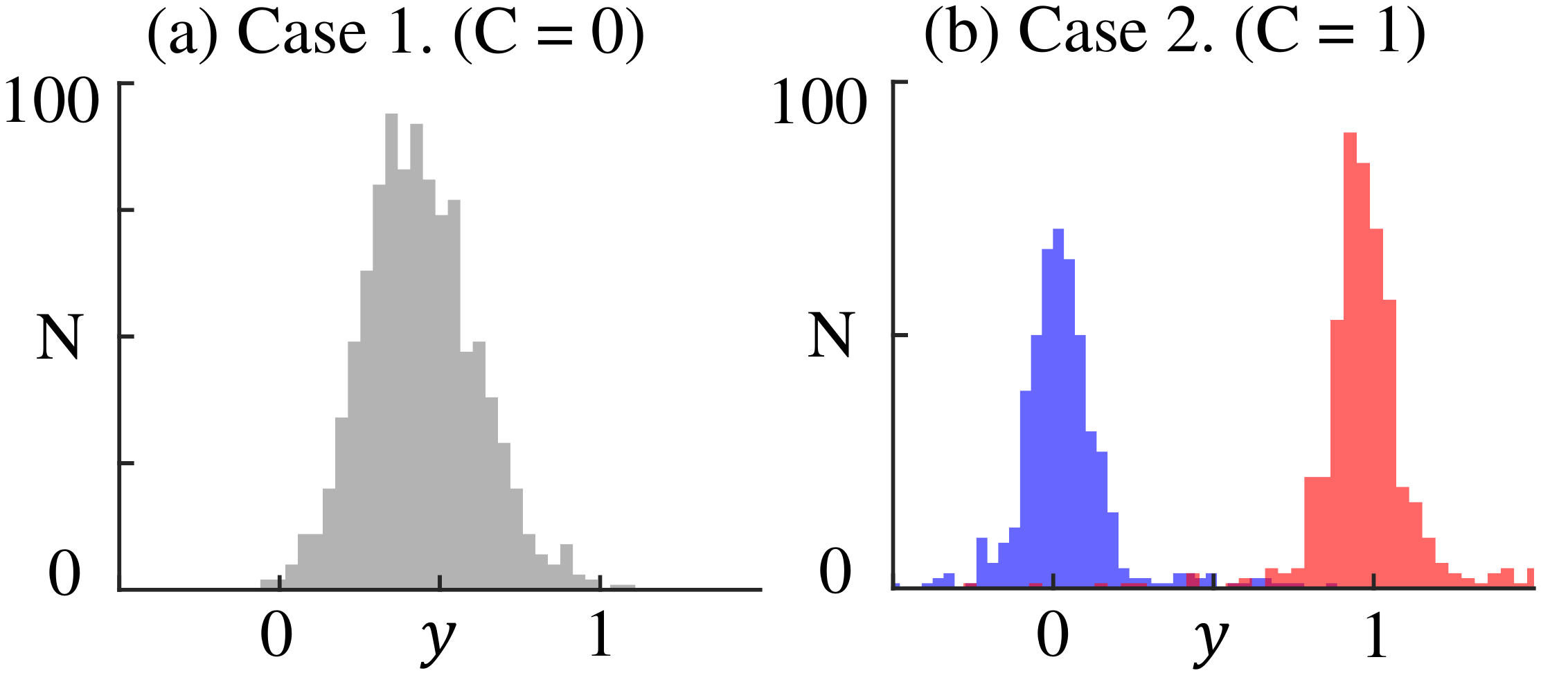} 
    \caption{The number of input child systems~N vs. the net's output $y$ evaluated on the test dataset. Panel (a) (panel (b)) corresponds to parent band insulator 1. (2.) from Fig. \ref{fig7}. In panel (b) we illustrate the data corresponding to the trivial (nontrivial) ensemble in blue (red).}
     \label{fig_C_10}\label{fig:2D_training}
\end{figure}

In Fig.\ \ref{fig:2D_training} we present the training outcome of our protocol corresponding to cases 1.\ and 2.\ from Fig.\ \ref{fig:Chern} obtained after training for $1000$ epochs with learning rate $10^{-4}$ and $1000$ epochs with learning rate $10^{-5}$. As expected the network (un)successfully learned to separate the dataset of topologically nontrivial (trivial) child Bloch Hamiltonians from a trivial reference. Note that the plotted distributions over the network's output $y$ are considerably broader in this case than the analogous result in 1d, cf.\ Fig.\ \ref{fig:random_class1D}. We expect that this is related to the more complex nature of the Chern number compared to the the winding number and therefore it is harder to train a network to represent it. Nevertheless, the obtained classification accuracies are very close to optimal: $51 \%$ and $98 \%$ for the cases of trivial and nontrivial ensembles, respectively. We anticipate that the distributions can be made sharper by increasing the computational resources used for the training.

\begin{figure} \centering
    \includegraphics[width=8.5cm,angle=0]{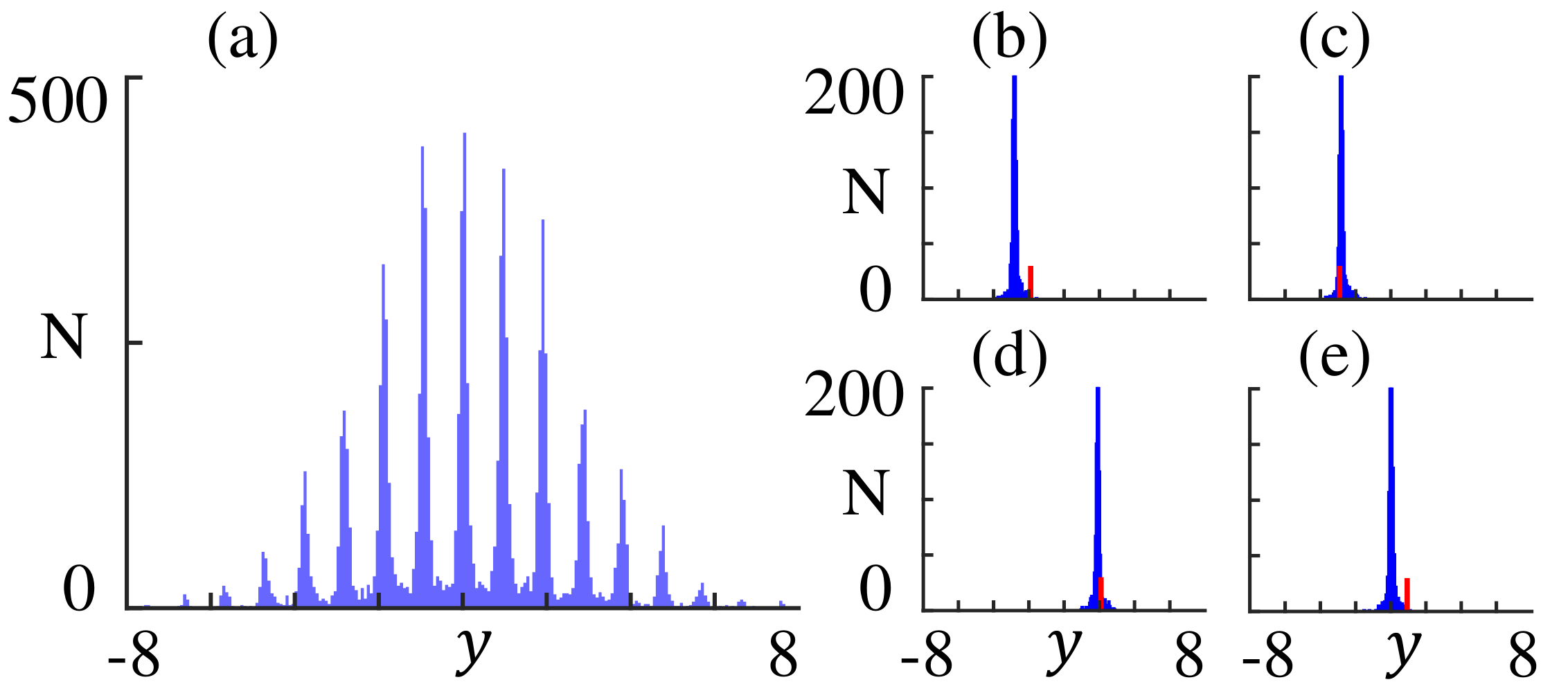}
    \caption{ The number of input child systems N vs.  network's output $y$ evaluated on (a) $10^4$ randomly selected Bloch Hamiltonians, (b) - (e) ensembles of $10^3$ topologically equivalent Bloch Hamiltonians generated from four distinct random parent systems (red marks).}
     \label{fig_C_random_1}\label{fig:random_class2D}
\end{figure}

In Fig.\ \ref{fig:random_class2D}a we present the output from $10^4$ random 2d band insulators using the network trained by the $C=1$ ensemble. The data is generated by randomly selecting 3d unit vectors on $10$ by $10$ momentum grid, subsequently expanded to $20\times 20$ by linear interpolation. As for the 1d case we find a classification of data according to topological sectors. We then select four random parent Bloch Hamiltonians, out of each of them create $10^3$ topologically equivalent child systems by following the deformation protocol, and push the obtained data through the trained network. The topologically equivalent systems are anticipated to group around the same output which is confirmed in Figs.~\ref{fig_C_random_1}b-e. Strikingly, even in the case that the network gives an uncertain or even incorrect classification of a particular object, such as in Figs.\ \ref{fig:random_class2D}b and e, the corresponding ensemble of children is peaked at the correct index of the object. In this way, although requiring extra computational effort, we can classify any object with a fidelity arbitrarily close to $100 \%$.     

In Appendix A we show that also for the case of 2d band insulators we can make the protocol completely unsupervised, training on an ensemble derived from a random parent object with unknown Chern number. The classification is in terms of an integer fraction which can be transformed to an integer index by a simple rescaling, cf.~Fig.\ \ref{fig:random_random_2D}.  

\begin{figure} \centering
    \includegraphics[width=8.5cm,angle=0]{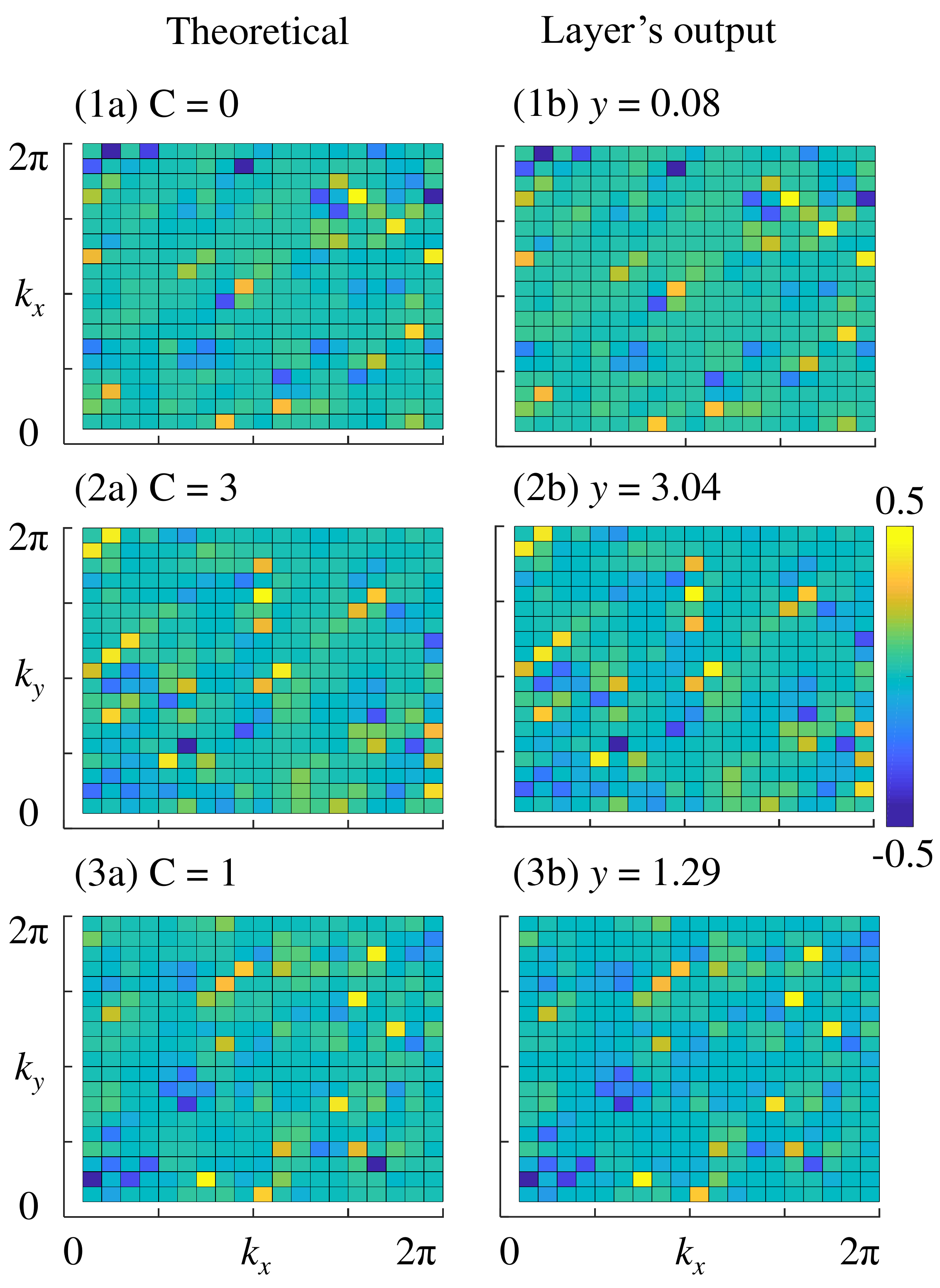}
    \caption{Samples of feature maps from the last convolutional layer  of the network in Fig.\ \ref{fig:2D_network} (right), compared to the explicitly calculated Berry curvature in discretized momentum space \cite{Zhang2} (left). These maps are summed up to give the network output and the exact Chern number respectively. This demonstrates that network has learnt to calculate the relevant local quantity.}
     \label{fig9_3}\label{fig:chern_local}
\end{figure}

In Fig.\ \ref{fig:chern_local} we compare the output of the last feature map, for three examples with random Hamilonians, compared to an explicit calculation of the discrete Berry curvature\cite{Zhang2}. Similarly to the 1d case we find that the network has learned a representation which is very close to the standard analytic representation. Once again, we attribute this to the fully convolutional, plus summation layer, structure of the network, which seems particularly well suited for the task.

\section{Summary \label{sec:summary}}
We have developed a neural-network-based unsupervised protocol for classifying topological phases and calculating their topological index. 
The algorithm is based on the core property that topology is robust to continuous deformations. 
Thus, it is not necessary to be able to calculate the index for producing ensembles of topologically equivalent objects, we just have to ensure that they can be continuously transformed into each other. 
We show how to derive such ensembles for the case of Bloch Hamiltonians of 2-band insulators in 1d and 2d and use these to calculate topological indices by training a purposefully designed convolutional neural network. Training the network on an ensemble of Hamiltonians with a priori unknown topological index, still allows the classification of an arbitrary object in the class. In addition, by classifying ensembles of topologically equivalent objects rather than individual objects we can achieve $100 \%$ classification accuracy.

In contrast to other ML studies of topological properties\cite{Zhang, Zhang2, Carvalho, Caio, Ming, Mano, Wu2}, our method does not require any training on labeled data with known topological index. The proposed approach, of data augmentation by continuous deformations, is in principle universal and it will be interesting to explore its potential for topological classification in more exotic symmetry classes (e.g. crystalline~TIs \cite{Fu2016}), in  higher dimensions, under time-periodic external drives\cite{Kitagawa, Lindner}, and/or in the presence of interactions~\cite{Ryu, Wen}. 

\section*{ACKNOWLEDGMENTS}

We thank Evert van Nieuwenburg for  discussions. This work was supported by the Swedish Research Council through Grant No. 621-2014-5972. \\

\appendix

\section{Training with unknown Chern number}

Here we show results for training the network with an ensemble dervived from a trivial reference Hamiltonian ($C=0$), labeled '0', and an ensemble derived from a parent Hamiltonian with unknown Chern number, labeled '1'.  In this way there is no bias whatsoever in choosing the parent used for the training. The ensemble of unknown index is derived from the random parent in the fashion described in Sec.\ \ref{sec:2D}. The network's architecture, the optimization method, and all the hyper parameters are also the same. 

\begin{figure} \centering
    \includegraphics[width=8.5cm,angle=0]{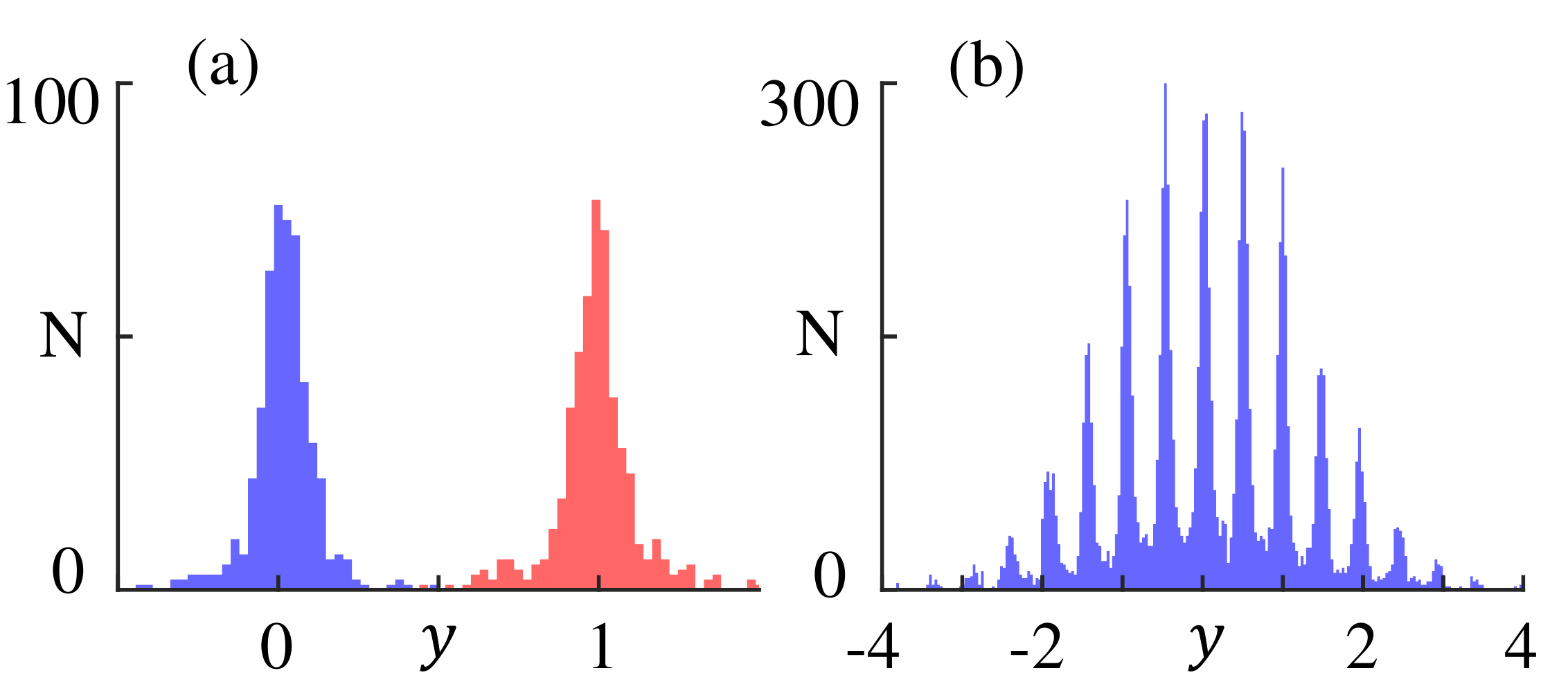}
    \caption{The outcome of our protocol performed on a randomly selected 2d parent system:  The number of input systems N vs.  the network's output $y$ corresponding to a) the test set of $10^3$ Bloch Hamiltonians and b) a set of $10^4$ randomly chosen 2d band insulators. In panel (b) the data corresponding to the trivial reference (randomly selected parent 2d band insulator) is illustrated in blue (red).}
     \label{fig_C2}\label{fig:random_random_2D}
\end{figure}

In Fig.~\ref{fig_C2} we depict the results for one example. From Fig.~\ref{fig_C2}a we note that the two datasets were successfully separated by the network indicating that the chosen parent system is topologically nontrivial. Moreover, the trained network has been also tested on a dataset of $10^4$ randomly generated Bloch Hamiltonians, Fig. ~\ref{fig_C2}b. In agreement with our anticipation all the outputs grouped at certain values. Similar to the 1d case (Sec.\ \ref{unknown_wind}) the output peaks do not position just at integers anymore; in fact they are half integer, and we can conclude that the parent system corresponds to Chern number $C = 2$.

\begin{figure} \centering
    \includegraphics[width=6.5cm,angle=0]{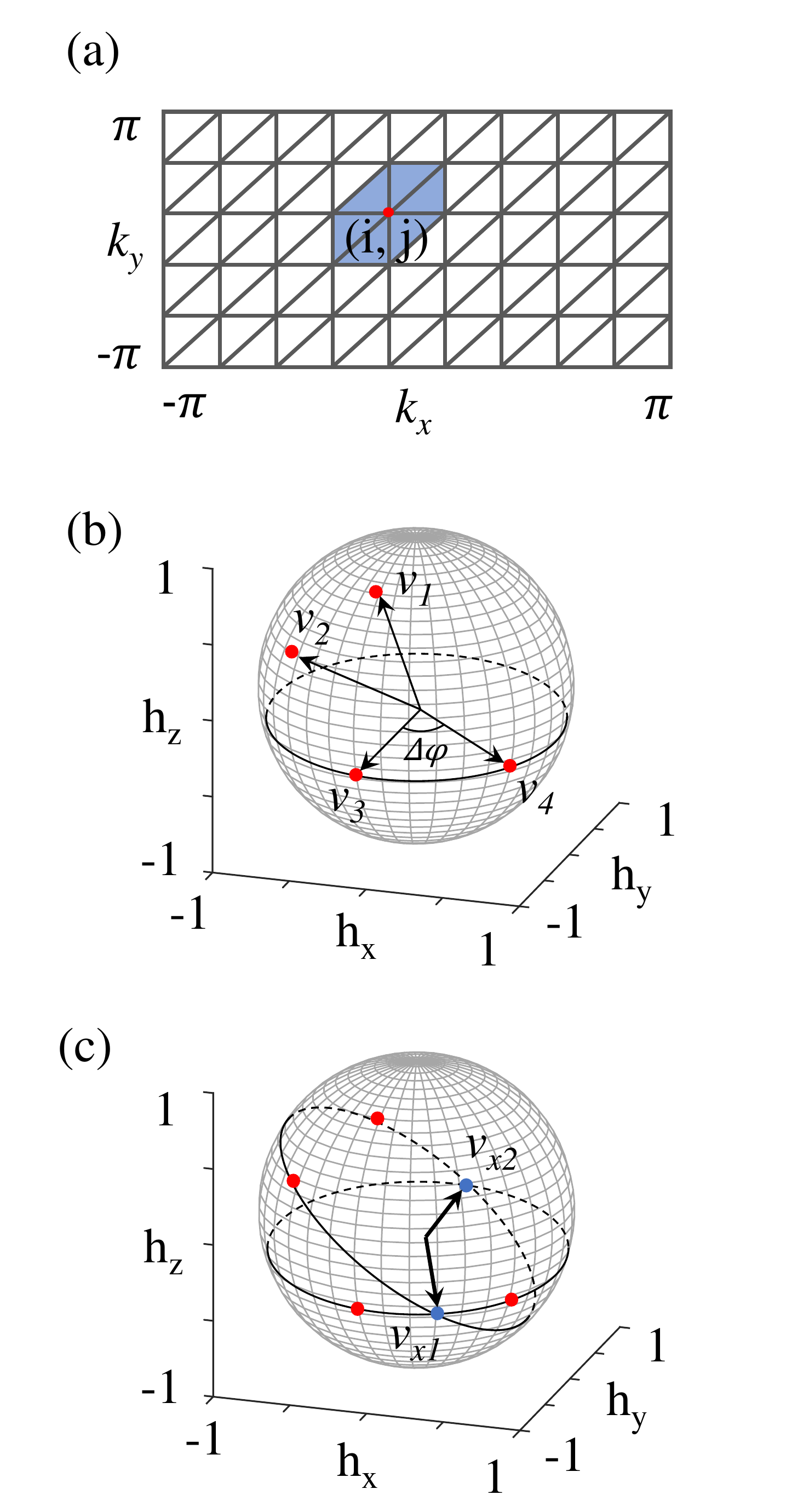}
    \caption{a) Six triangles on the momentum grid that get affected by a rotation of the vector at site $(i, j)$. b) Three vectors $v_1$, $v_2$, $v_3$ defining the corresponding spherical triangle, with $v_3$ to be rotated  around z axis by an angle $\Delta \phi$ to $v_4$. c) Vectors $v_{x1}$ and $v_{x2}$ being at the intersection of the equatorial planes $(v_1, v_2)$ and $(v_3, v_4)$.}
     \label{cont_criteria_abc} 
\end{figure}

\section{Continuous deformations of discretized Chern insulators\label{App:2d_transform}}

Here we develop a continuity condition for random transformations performed on 2d topological band insulators in symmetry class A. We consider a 2-band Bloch Hamiltonian $H(k_x, k_y)$ defined on some grid of $N_x \times N_y$ sites, interpolate it with the procedure described in Sec.\ \ref{sec:2D_training}, and then track how the corresponding interpolated system transforms under certain type of transformations, to determine whether a discontinuous change could occur. In Sec.\ \ref{sec:2D_training} the topological equivalence classes are surveyed using multiple basic transformations performed by rotating $H(k_x, k_y)$ at one randomly chosen momentum site by a random angle around z axis. Here we develop a continuity condition for any such deformation and in the numeric implementation of our protocol perform only the deformations that do not violate this criterion.

Assume that we aim to rotate the Bloch Hamiltonian $H(k_x, k_y)$ at momentum site $(i, j)$ by some angle $\Delta \phi$ around z axis. The deformation may cause a discontinuous change only at 6 triangles sharing site $(i, j)$, Fig.~\ref{cont_criteria_abc}a, and we need to independently examine each of them to ensure that the interpolated system has changed continuously. Consider a spherical triangle corresponding to one of those 6 triangles and denote the vectors corresponding to its vertices by $v_1$, $v_2$, and $v_3$ with the latter denoting the vector that has been chosen to be rotated by $\Delta \phi$ to vector $v_4$, Fig.~\ref{cont_criteria_abc}b. Recall that the interpolation is done through choosing the spherical triangle corresponding to the smallest area, cf. Sec.\ \ref{sec:2D_training}. It follows that the deformation is certainly continuous if the vector $v_3$ did not cross the equatorial plane defined by $v_1$ and $v_2$ during the deformation and we allow such deformations. If the vertex $v_3$ has crossed that plane, at point $v_x$, then it is discontinuous if the triangle defined by the vectors $v_1$, $v_2$, and $v_x$ is acute. (Corresponding to the spherical triangle area going through $2\pi$.)  Thus, we forbid any deformations that cause the vector $v_3$ to cross the equatorial plane defined by $v_1$ and $v_2$ at $v_x$ such that the resulting triangle defined by $v_1$, $v_2$, and $v_x$ is acute, i.e. all its angles are smaller than $\pi/2$.

We now develop a straightforward analytic solution describing this criteria. First we find $v_x$ by looking for a vector normal to $n = v_1 \times v_2$ and having the same azimuthal angle  as $v_3$: $\theta_x = \theta_3$. In general there can be up to two solutions for $v_x$, Fig.~\ref{cont_criteria_abc}c. They are solutions to the following equation to be solved for the polar angle $\phi_x$ corresponding to $v_x$:
\begin{align} 
\begin{split} 
0 &= n_x   \sin(\theta_x)\sin(\phi_x) + n_y  \sin(\theta_x)\cos(\phi_x) + n_z \cos(\theta_x)  \\
 &=A  \sin(\phi_x) + B  \cos(\phi_x) + C,\\
\end{split}
\label{eq:T_cond}
\end{align}
where $A = n_x \sin(\theta_x)$, $B = n_y \sin(\theta_x)$, and $C = n_z \cos(\theta_x)$ with $n = (n_x, n_y, n_z) = v_1 \times v_2$ and $\theta_x = \theta_3$. The solution can be found through solving $ax^2 + bx + c = 0$ with $x = \cos(\phi_x)$, $a = A^2 + B^2$, $b = 2AC$, $c = C^2 - B^2$, and $\sin(\phi_x) = - (Ax + C)/B$. There will be up to two $\phi_x$ satisfying this equation. For each resulting $\phi_x$ we first check whether a triangle defined by $v_1$, $v_2$, and $v_x$ is acute-angled or not. If not then we allow this deformation. If it is acute-angled for at least one $v_x$ then we explicitly check whether $v_3$ crosses it during the deformation: we add $\Delta \phi$ to $\phi_3$ and verify if the polar angle reaches $\phi_x$. If yes, then we forbid this deformation to be performed, otherwise, it is allowed. We repeat this analysis for each of the 6 triangles sharing the vector at site $(i,j)$ and only implement the deformations that are allowed (continuous) for all of them. In this way the implemented deformations are ensured to preserve the equivalence class. 

\end{document}